# Kant: An Efficient Unified Scheduling System for Large-Scale AI Clusters

LINGLING ZENG*, GEN ZHANG, JIALIN PENG, XIANG XU, YUAN XU, LIJUN MA

ZTE Corporation

**Abstract**

As AI cluster sizes continue to expand and the demand for large-language-model (LLM) training and inference workloads grows rapidly, traditional scheduling systems face significant challenges in balancing resource utilization, scheduling efficiency, and service quality. This paper presents and evaluates **Kant**: an efficient unified scheduling platform designed for large-scale AI container clusters, supporting the co-scheduling of both training and inference jobs.

Based on the practical implementation of the Kant system, we systematically define a set of key evaluation metrics for AI clusters, including GPU Allocation Ratio (GAR), Scheduling Occupancy Rate (SOR), GPU Node Fragmentation Ratio (GFR), Job Waiting Time Distribution (JWTD), and Job Training Time Estimation Distribution (JTTED), providing a foundation for quantitative performance analysis.

Experimental results demonstrate that Kant achieves exceptional performance in clusters ranging from hundreds to tens of thousands of GPUs. By leveraging scheduling strategies such as Backfill and Enhanced Binpack (E-Binpack), the system significantly improves resource utilization and scheduling efficiency, while effectively reducing resource fragmentation and communication overhead in distributed training. The system has been deployed in multiple AI data center clusters, where it stably supports large-scale intelligent computing workloads.

This work provides a practical engineering approach for building high-performance, highly available, AI-native scheduling infrastructure.

## 1 INTRODUCTION

In recent years, with the widespread adoption of LLM, the demand for AI computing has surged dramatically. To support the growing need for large-scale AI training and inference tasks, AI cluster scales continue to expand—from hundreds to thousands, tens of thousands, and even "hundred-thousand-GPU" systems—some of which are already deployed in production environments.

Meanwhile, AI clusters exhibit highly diverse characteristics: GPU device models are becoming increasingly heterogeneous, resource demands vary significantly across tenants, and job types range from large-scale distributed training to online inference services, spanning from single-GPU to multi-thousand-GPU scales. This diversity imposes higher requirements on the adaptability, flexibility, and scheduling efficiency of cluster resource management systems.

---

* Corresponding author.

An efficient resource scheduling mechanism is a critical component for ensuring high-performance AI cluster operations. It not only affects GPU utilization but also directly determines job execution performance, service quality (Quality of Service, QoS), and overall system throughput. Therefore, designing a scheduling system capable of adapting to large-scale, heterogeneous, and highly dynamic AI workloads holds significant theoretical and practical value.

This paper presents **Kant**—a unified scheduling platform for AI clusters built on Kubernetes [3]. Tailored to the characteristics of AI workloads, Kant features an efficient scheduling architecture and resource allocation strategies, supports co-scheduling of training and inference jobs, and aims to improve GPU utilization, ensure service quality, and reduce job waiting times. We also systematically define key evaluation metrics such as GAR, SOR, GFR, JWTD and JTTED, providing a foundation for quantitative performance analysis.

Experimental results show that Kant achieves excellent performance across multiple key indicators, significantly improving scheduling efficiency, resource utilization, and service quality, thereby effectively addressing the scheduling challenges of large-scale AI clusters.

Currently, the system has been deployed in multiple AI container clusters and is stably supporting large-scale intelligent computing workloads.

## 2 MOTIVATION AND CHALLENGES

Current AI clusters exhibit the following key characteristics:

- Scalability: GPU clusters are expanding from hundreds to thousands, tens of thousands, and even hundred-thousand-GPU scales, providing a powerful computing foundation for complex AI tasks [1].
- GPU Heterogeneity: Modern AI clusters often deploy multiple GPU models, forming heterogeneous resource pools. These models differ significantly in computational performance, memory bandwidth, and communication capabilities, posing technical challenges for resource scheduling.
- Tenant Diversity: In multi-tenant environments, resource demands vary widely among tenants. GPUs are typically allocated in non-shared mode, though shared mechanisms may be used under resource pressure, increasing scheduling complexity.
- Wide Distribution of Job Sizes: In clusters with tens of thousands of GPUs [1], over 90% of jobs use fewer than 8 GPUs, yet their cumulative GPU-time accounts for less than 10% of the total. In contrast, jobs requiring 256 GPUs or more, though fewer in number, consume over half of the total GPU computing time. For example, large-scale training tasks such as Llama3 may involve tens of thousands of GPUs [2]. This disparity presents significant challenges for scheduling strategy design.
- Diverse Task Types: The system must support various task types, including:
  - LLM Distributed Training: Emphasizing training efficiency;
  - Inference Services: Focusing on low latency and high availability;
  - Development and Debugging Tasks: Prioritizing flexibility and rapid response.

These tasks differ significantly in resource requirements, scheduling behavior, and QoS objectives.

These characteristics pose the following key challenges to resource scheduling systems:

1. How to achieve fair scheduling and resource isolation in multi-tenant, heterogeneous GPU clusters;
2. How to enable efficient and rational job scheduling in large-scale, heterogeneous, and dynamic environments;

3. How to optimize GPU resource allocation to reduce idle time and fragmentation, thereby improving overall utilization;
4. How to meet the scheduling needs of diverse task types and priorities while ensuring SLA compliance;
5. How to enhance system throughput while optimizing user experience.

Depending on the application scenario, current mainstream AI scheduling systems fall into two categories, each with limitations:

- HPC schedulers (e.g., SLURM [5]): Originally designed for scientific computing, they offer mature resource management and high stability. However, their support for container orchestration and elastic scheduling is limited, making them less suitable for modern AI workloads.
- Containerized schedulers (e.g., Kubernetes native scheduler, Volcano [4]): Designed for cloud-native environments, they provide strong flexibility and scalability for AI task management. However, they often suffer from low resource utilization, high scheduling latency, and increased risk of SLA violations when handling large-scale, heterogeneous AI workloads.

Given the scale and diversity of modern AI clusters, this paper proposes and implements Kant—a unified scheduling system aimed at achieving multi-tenant fairness, improving resource utilization, reducing scheduling latency, enhancing SLA assurance, and optimizing user experience.

## 3 KANT SYSTEM DESIGN

To address the core scheduling challenges in AI clusters, Kant employs a unified architecture focused on multi-tenant fairness, high resource utilization, and strong service quality assurance. Key components include:

1. **Multi-tenant Fair Scheduling and Resource Isolation**
   Through mechanisms such as admission control (3.2.1), queuing control (3.2.2), and preemption control (3.2.3), Kant ensures efficient job management and resource fairness.
2. **Efficient Scheduling in Large-Scale, Heterogeneous, and Dynamic Environments**
   By introducing hierarchical scheduling with node grouping, cluster partitioning for heterogeneity, and scheduler memory access optimization (3.4), Kant significantly improves scheduling throughput and system scalability.
3. **Improving GPU Resource Utilization and Reducing Resource Idle Time**
   Leveraging strategies such as Backfill queuing (3.2.2), fine-grained scheduling (3.3.1), Gang scheduling (3.3.2), and enhanced Binpack scheduling (3.3.3), Kant maximizes GPU utilization.
4. **Meeting SLA Requirements for Different Priorities and Task Types**
   Based on Backfill queuing (3.2.2) and enhanced Spread scheduling (3.3.4), Kant enables fast scheduling of high-priority jobs and controlled delay for low-priority ones, ensuring high availability.
5. **Optimizing User Experience and System Throughput**
   Through enhanced Binpack scheduling (3.3.3) and topology-aware scheduling (3.3.5), Kant improves resource utilization while balancing job execution efficiency and communication performance.

The following sections will first present the overall architecture of Kant, followed by detailed explanations of its core modules and key technical implementations.

## 3.1 System Architecture

The Kant system is a distributed, unified scheduling platform built on Kubernetes, specifically designed for AI workloads. It aims to address the complex challenges of resource scheduling in large-scale, highly heterogeneous, and multi-tenant AI clusters.

Centered on high-performance scheduling and fine-grained resource management, Kant is designed to improve GPU resource utilization, optimize job execution performance, and ensure service quality (QoS), thereby providing stable and efficient scheduling capabilities in dynamic AI workload environments.

Kant is developed based on the Kubernetes scheduling framework and consists of two core components: QSCH (Queue-based Scheduler) and RSCH (Resource-aware Scheduler). These components handle job queue management and resource allocation decisions, respectively, jointly enabling a full scheduling pipeline — from AI job submission to resource allocation—as illustrated in Figure 1.

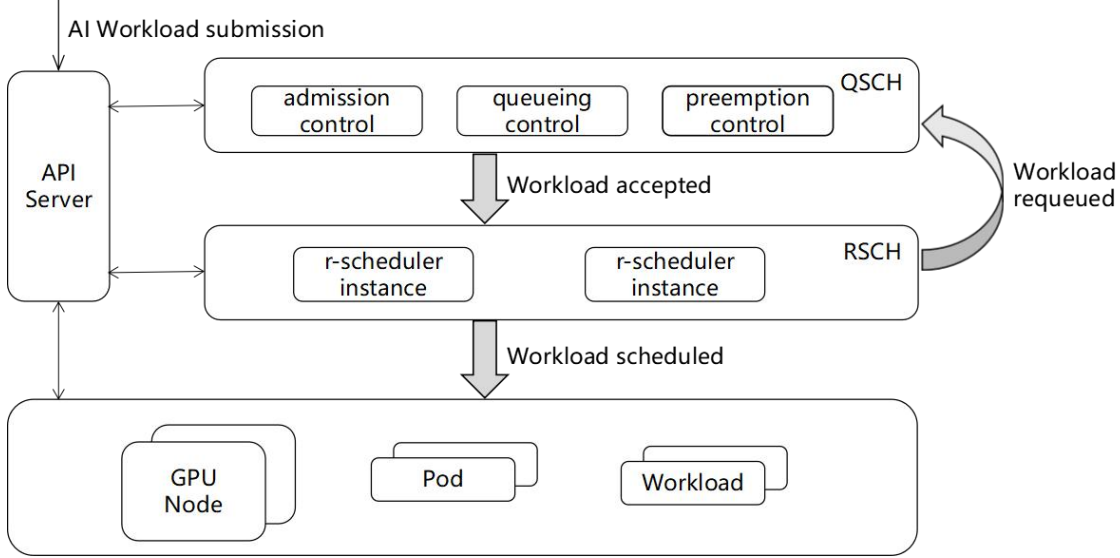

Figure 1: Kant system architecture.

1. **QSCH (Queue-based Scheduler)**
   Manages job queuing, admission control, and preemption strategies. It is the core module for achieving fair scheduling across multiple tenants and task types.
2. **RSCH (Resource-aware Scheduler)**
   Focuses on fine-grained resource allocation and scheduling decisions, supporting various scheduling strategies for diverse AI tasks. It supports multi-instance deployment to achieve higher scheduling throughput in large-scale clusters.

By decoupling queue control and resource allocation, the collaborative design of QSCH and RSCH enhances system scalability and scheduling efficiency. This architecture considers multiple dimensions—including resource utilization, scheduling latency, QoS, and user experience — and further improves overall performance through the following key technical enhancements:

1. Finer-grained resource isolation and fairness guarantees in multi-tenant, heterogeneous environments;
2. Topology-aware scheduling to optimize communication performance for distributed jobs;
3. Enhanced Binpack (E-Binpack) and Enhanced Spread (E-Spread) scheduling strategies for balancing resource utilization and high availability;
4. Support for high-concurrency scheduling in large-scale cluster environments.

**3.2 QSCH Queue-based Scheduler**

QSCH is the core module in Kant responsible for job queuing and admission control. Its main functions include:

- Admission Control
- Queueing Control
- Preemption Control
- Requeueing Mechanism

These mechanisms ensure that jobs pass resource availability checks before entering scheduling and have fallback paths upon scheduling failure.

*3.2.1 Admission Control*

QSCH selects the appropriate admission granularity based on job type:

- **Gang-type jobs** (e.g., distributed training): Use **job level** admission, where the entire job proceeds only when all required resources are available.
- **Non-gang-type jobs** (e.g., inference services inference services with traditional non-distributed architectures): Employ **pod level** admission, where each pod is independently evaluated and, upon resource availability, admitted into the subsequent scheduling pipeline.

QSCH employs a two-tier admission mechanism: jobs undergo static quota admission followed by dynamic resource admission, proceeding only after passing both checks.

**Static Quota Admission**

AI clusters fall into two categories:

- **Homogeneous clusters**: Nodes use the same GPU model.
- **Heterogeneous clusters**: Nodes have varying GPU models.

To address resource incomparability due to GPU performance differences, QSCH groups nodes by GPU model into **GPU Type-based Node Pools** and supports tenant-specific quotas for each GPU model.

Quota management supports two modes:

- **Shared Mode**: Tenants can borrow unused quotas from others to improve utilization.
- **Isolated Mode**: Tenants are strictly limited to their own quotas, ensuring strong isolation.

If a tenant has sufficient available quota for the requested GPU model, the job is admitted; otherwise, it enters the queue to wait for resource release.

**Dynamic Resource Admission**

Even if a job passes static quota admission, it cannot be scheduled without sufficient free GPU resources. To address this, QSCH performs dynamic admission based on real-time cluster status:

- **Resource Readiness Check**: Prevents invalid scheduling attempts. If current resources are insufficient, the job fails admission and enters the waiting queue; otherwise, it proceeds to scheduling.

For homogeneous jobs (single GPU model), QSCH checks availability in the corresponding node pool. For heterogeneous jobs (multiple GPU models), it supports cross-pool joint admission to meet complex resource requirements.

### 3.2.2 Queueing Control

As a high-cost, limited resource, GPU is managed at the cluster level. To ensure fairness in multi-tenant environments, QSCH maintains separate job queues per tenant. When a user submits a job, it first enters the tenant-specific queue. Only after passing static quota admission does the job enter the global scheduling process, where Kant performs dynamic resource admission and allocation based on real-time status.

Key job ordering criteria include:

- Job Priority
- Submission Time
- Job Size (used as a tiebreaker when priorities are equal)

QSCH supports multiple queuing strategies, among which the Backfill strategy balances resource utilization and job response latency.

Table 1: Queueing Policy

| Queueing Policy | Working Mechanism | Applicable Scenarios and Advantages | Main Limitations |
|---|---|---|---|
| Strict FIFO | Jobs are scheduled strictly in the order of arrival. If the job at the front cannot be scheduled due to insufficient resources, all subsequent jobs must wait. | Ensures fairness and execution order; simple to implement. | May lead to resource fragmentation and reduced utilization due to head-of-line blocking. |
| Best-Effort FIFO | When the front job cannot be scheduled, smaller subsequent jobs may bypass it if sufficient resources are available, thereby improving throughput by utilizing otherwise idle resources. | Suitable for resource-constrained environments where high throughput is prioritized over strict fairness; reduces latency for small jobs. | Risks indefinite postponement (starvation) of large jobs, compromising fairness. |
| Backfill | If the front job cannot be scheduled immediately, smaller jobs are allowed to use available resources temporarily. Once the front job's waiting time exceeds a threshold, the system preempts the backfilled jobs to allocate resources to it. | Balances high resource utilization with fairness; prevents both resource waste and long-job starvation; effective for mixed-size workloads. | Increased implementation complexity and overhead due to preemption. |

### 3.2.3 Preemption Control

To improve resource utilization and ensure timely execution of high-priority jobs, QSCH supports the following preemption mechanisms:

- Priority Preemption: High-priority jobs can preempt resources from lower-priority ones.
- Quota Reclamation Preemption: The original quota owner can reclaim loaned resources via preemption.
- Backfill Preemption: In the Backfill strategy, if the head job waits too long due to insufficient resources, it can acquire them via preemption after a timeout.

For gang-type jobs, preemption occurs at the job level; for non-gang-type jobs, it is performed at the pod level.

Note that preemption involves resource release and reallocation, which may incur system overhead. In some cases, it can trigger cascading adjustments, potentially causing low-priority jobs to starve due to frequent preemption. Therefore, QSCH employs a conservative preemption policy, triggering preemption only under strict conditions to minimize negative impacts on scheduling stability.

### 3.2.4 Requeueing Mechanism

To enable automatic recovery from scheduling failures, Kant supports a requeueing mechanism:

- When scheduling fails due to resource insufficiency or conflicts, QSCH deletes the associated Pods upon detection and triggers the workload to re-enter the scheduling queue, restarting the process.

This mechanism avoids scheduling deadlocks and resource waste, enhances system robustness, and improves overall scheduling efficiency.

*3.2.5 Summary*

QSCH achieves efficient scheduling of complex jobs in large-scale AI clusters through a multi-level admission control mechanism (static quotas and dynamic resource assessment), flexible queuing and preemption strategies, and a robust requeueing mechanism. This design strikes a balance between resource utilization, job response efficiency, and multi-tenant fairness, laying a solid foundation for Kant's overall scheduling capabilities.

**3.3 RSCH Resource-aware Scheduler**

RSCH is the core component in the Kant system responsible for GPU resource allocation. Its primary objective is to accurately allocate GPU devices and associated resources based on job requests and actual node resource conditions, ensuring efficient task execution and compliance with service quality (QoS) requirements.

*3.3.1 Fine-Grained GPU Scheduling*

RSCH supports fine-grained, device-level scheduling. It not only evaluates GPU availability at the node level but also precisely assigns specific GPU devices and their topologically associated high-performance components (e.g., RDMA NIC) to Pods.

This mechanism continuously monitors the health status of GPUs and connected hardware, dynamically allocating optimal device resources based on topology awareness and health information.

It is particularly suitable for:

- Small-scale AI jobs that request fractional or non-multiple GPU resources per node, enabling efficient utilization of fragmented resources;
- Large-scale AI jobs sensitive to device health and communication topology, significantly improving resource utilization and execution stability.

*3.3.2 Gang Scheduling*

For distributed AI jobs, RSCH implements All-or-Nothing scheduling (also known as Gang scheduling), where scheduling proceeds only when all required resources for the target Pods can be simultaneously satisfied; otherwise, the job remains in a pending state.

This mechanism prevents partial scheduling—where some Pods are scheduled while others cannot start—reducing invalid scheduling attempts and resource waste, and ensuring execution consistency and efficiency for distributed AI workloads.

*3.3.3 Binpack Scheduling Enhancements*

RSCH provides a Binpack scheduling strategy that prioritizes placing Pods on nodes with existing GPU allocations, aiming to fill partially utilized nodes and minimize the activation of new nodes.

This strategy is ideal for training clusters or resource-constrained environments, offering the following benefits:

- Reduces GPU node fragmentation and improves resource consolidation;
- Concentrates workloads, reserving full-node resources for large-scale AI jobs and reducing job blocking;
- Enables dynamic power-down of long-idle nodes, reducing overall cluster energy consumption.

**Enhanced Binpack (E-Binpack)**

The Kant system introduces E-Binpack in RSCH to further optimize job performance and reduce system-wide communication overhead.

- Node-level E-Binpack: Co-locate multiple Pod replicas of the same job on the same node to minimize cross-node communication.
- LeafGroup-level E-Binpack: Consolidate small AI jobs within the same NodeNetGroup to reduce inter-group traffic, while reserving entire group resources for large jobs.

Additionally, the Kant system plans to introduce a periodic fragmentation reorganization mechanism that consolidates scattered resources via rescheduling, further improving utilization.

### 3.3.4 Spread Scheduling Enhancements

RSCH provides a Spread scheduling strategy that distributes multiple Pod replicas of AI inference services across different nodes, achieving balanced GPU utilization.

This strategy is primarily used in non-distributed inference scenarios and offers:

- Enhanced fault tolerance;
- Improved system high availability and robustness.

**Enhanced Spread (E-Spread)**

With the rise of large-scale inference models, cross-node distributed architectures—such as the 64-way Expert Parallelism (EP) spanning eight nodes in DeepSeek-V3 [6] and the KVCache-centric disaggregated serving system Mooncake [7]—are becoming increasingly common. These workloads require multiple entire nodes to host expert replicas or pipeline stages, making node fragmentation a critical issue. Traditional spread-based placement strategies scatter free GPUs across the cluster, leaving no single node with sufficient continuous resources and thus degrading acquisition success for large jobs.

To address this, Kant introduces E-Spread—**Inference Dedicated Zone Scheduling**. Using an 8-GPU node as an example:

- A subset of GPU nodes is designated as an inference dedicated zone;
- Inference Pods requiring fewer than 8 GPUs are prioritized for Spread scheduling within this zone;
- Remaining replicas are scheduled using E-Binpack in the general resource pool.

Advantages of E-Spread include:

- Supports highly available deployment of small-scale inference services;
- Limits resource dispersion of inference workloads to a defined area;
- Preserves full-node resources for large-scale distributed inference tasks;

- Compatible with training-inference mixed deployment, improving overall cluster utilization.

*3.3.5 Topology-Aware Scheduling*

In AI clusters, multiple interconnection paths exist between GPUs, with significant performance differences. RSCH employs a topology-aware scheduling mechanism to intelligently place Pods based on communication link quality, maximizing job performance.

**Intra-Node GPU Topology**

Within a single node, GPUs are interconnected via NVLink, PCIe, and NUMA domains, with communication bandwidth decreasing in that order.

For multi-GPU jobs, RSCH selects GPU combinations with optimal interconnect performance and pairs them with the RDMA NIC offering the best communication path, enhancing overall efficiency.

This mechanism applies to all AI clusters and is especially critical for communication-intensive workloads such as distributed training.

**Inter-Node GPU Topology Scale-Out**

The RDMA network in AI clusters typically uses a Scale-Out Interconnect architecture, with a hierarchical structure comprising:

- Access layer (Leaf)
- Aggregation layer (Spine)
- Core layer (Superspine)

RSCH selects scheduling locations based on communication quality, with the following preference:

1. Nodes within the same Leaf layer
2. Nodes within the same Spine layer
3. Nodes within the same Superspine layer

Lower-tier connections generally have lower latency; thus, RSCH prioritizes them to reduce communication overhead and improve job efficiency.

**Inter-Node GPU Topology Scale-Up**

In hyper-node scenarios, a Scale-Up Interconnect architecture exists, forming HBD (Hyper Bandwidth Domain) domains. Within an HBD, all node GPUs are interconnected at high speed, supporting EP (Expert Pipeline) or TP (Tensor Pipeline) communication patterns.

For large AI jobs with EP communication needs, RSCH schedules them at the HBD granularity.

*3.3.6 Summary*

RSCH integrates fine-grained device-level scheduling, Gang scheduling, E-Binpack and E-Spread, and topology-aware scheduling. These mechanisms collectively address the diverse resource allocation needs of AI clusters — from training to inference, and from small to large-scale workloads.

Together, they significantly improve GPU utilization, job performance, and system stability, laying a solid foundation for a high-performance, low-latency, and highly reliable AI scheduling system in Kant.

### 3.4 Performance Optimization Mechanism for Scheduling

To address scheduling efficiency bottlenecks in large-scale AI clusters, the Kant system introduces several key performance optimization mechanisms, covering heterogeneous resource management, group-based scheduling, and memory operation optimization. These mechanisms work synergistically to significantly improve scheduling throughput, reduce job response latency, and enhance overall system stability.

#### 3.4.1 Splitting Heterogeneous Clusters

In GPU heterogeneous clusters, nodes are equipped with diverse GPU models that differ significantly in computing power and communication performance. To enhance scheduling efficiency, the Kant system divides the cluster into multiple "**GPU Type-based Node Pools**" based on GPU model.

During job scheduling, the scheduler searches only within node pools matching the requested GPU model, avoiding global traversal. This splitting strategy significantly reduces the scheduling search space, lowers computational overhead, and thereby improves scheduling efficiency.

#### 3.4.2 Hierarchical Scheduling with Node Grouping

To alleviate performance pressure on scheduling components in large-scale AI clusters, the Kant system introduces a hierarchical scheduling mechanism based on hardware topology, enabling efficient resource allocation through node grouping and a two-stage scheduling process.

**NodeNetGroup**

As described in Section 3.3.5, the typical Scale-Out hierarchical structure includes access (leaf), aggregation (spine), and core (superspine) layers of switches.

The Kant system abstracts each LeafGroup as a NodeNetGroup, which serves as the basic scheduling management unit and reflects the communication affinity of the underlying network topology.

**Two-Level Scheduling**

The scheduling process proceeds as follows:

1. Group-level preselection: The scheduler filters candidate NodeNetGroups (Selected Groups) based on their resource availability;
2. Node selection within group: Selects target nodes (Selected Nodes) within the chosen group for Pod placement.

This mechanism offers the following advantages:

- Significantly reduces the scheduling search scope, lowering scheduling complexity and improving efficiency;
- Supports integration with strategies such as Group-level E-Binpack to optimize job performance and resource consolidation;
- Enables parallel scheduling across groups, enhancing overall system throughput.

*3.4.3 Memory Optimization in Scheduling*

To ensure data consistency, schedulers typically fetch a snapshot of the cluster resource state from the cache before each scheduling cycle and create an independent copy via deep copying. However, in large-scale clusters, this deep copy operation becomes a key performance bottleneck.

To address this, RSCH introduces an incremental update mechanism that copies only the data portions modified since the last scheduling cycle. This significantly reduces redundant data copying and greatly lowers memory and CPU overhead in the scheduling component.

Experimental results show that after introducing the incremental update mechanism in a test cluster with 1,000 nodes, RSCH's CPU load decreased by more than 50%, and both the overall scheduling performance and stability of the Kant system were significantly improved.

*3.4.4 Summary*

Scheduling performance optimization mechanisms are essential to the efficient operation of the Kant system. Through key technologies such as heterogeneous cluster splitting, hierarchical group scheduling, and memory operation optimization, the Kant system effectively reduces scheduling complexity, improves resource utilization and job response efficiency, and enhances stability and scalability in large-scale AI clusters—providing solid support for its high-performance scheduling capabilities.

## 4 KEY PERFORMANCE METRICS FOR SCHEDULING

We have defined a set of quantitative evaluation metrics for AI cluster scheduling systems in the Kant system. These metrics, spanning multiple dimensions—including resource utilization, scheduling efficiency, and service quality—objectively reflect the comprehensive performance of the scheduling system under different load scenarios.

Excellent performance of an AI cluster scheduling system is typically reflected in the following aspects:

- High GPU resource allocation efficiency;
- Continuous and stable resource occupancy capability;
- Low resource fragmentation rate;
- Fast response to job requests;
- Reasonable training performance assurance.

Based on these requirements, we have built a multi-dimensional evaluation framework covering resource utilization, scheduling efficiency, and job performance, comprising the following five key metrics:

1. GPU Allocation Ratio (GAR)
2. Scheduling Occupation Ratio (SOR)
3. GPU Node Fragmentation Ratio (GFR)
4. Job Waiting Time Distribution (JWTD)
5. Job Training Time Estimation Distribution (JTTED)

The following sections define and explain each metric, supported by analysis of real scheduling scenarios, to facilitate subsequent experimental evaluation and strategy optimization.

## 4.1 GPU Allocation Ratio (GAR)

**Definition**

The GPU Allocation Ratio (GAR) measures the overall extent to which GPU resources in the cluster are allocated by the scheduling system. It is defined as:

$$GAR = \frac{N_{allocated}}{N_{total}}$$

where:

- $N_{allocated}$ denotes the number of GPUs currently occupied by AI workloads (including training and inference tasks);
- $N_{total}$ denotes the total number of available GPUs in the cluster.

**Explanation and Analysis**

GAR is one of the key metrics for evaluating resource utilization in scheduling systems. Ideally, under sufficient job demand and resource availability, GAR should approach 100%. However, in practice, GAR often deviates from this theoretical optimum due to various factors.

- **Main Influencing Factors**
    - **Insufficient Job Demand**: When the number of job requests is low, the system may not fully allocate available resources even if they are abundant, resulting in a low GAR. In such cases, a low GAR reflects underutilized demand rather than poor scheduling efficiency.
    - **Resource Fragmentation and Topological Constraints**: Even with idle GPUs present, they may be unusable due to resource fragmentation or topological requirements. For example, a job requiring four contiguous GPUs may fail to schedule if only scattered single-GPU nodes are available, thereby reducing GAR.
- **Recommendation for Comprehensive Evaluation** GAR alone should not be used as the sole indicator of scheduling efficiency. To provide a more holistic assessment of the system's resource utilization capability, we recommend combining GAR with the following auxiliary metrics:
    - **GPU Fragmentation Ratio (GFR)**: Measures the degree to which resource dispersion prevents large jobs from being scheduled. A lower GFR indicates stronger resource consolidation and higher allocatability.
    - **Job Waiting Time Distribution (JWTD)**: Reflects the latency between job submission and scheduling initiation. Shorter waiting times indicate faster scheduling response and higher efficiency.

Note: It is important to distinguish GAR from GPU Utilization Ratio, which is typically obtained via hardware monitoring tools such as DCGM (e.g., DCGM_FI_DEV_GPU_UTIL). This metric measures the actual computational intensity of GPU usage. The two metrics differ fundamentally in what they assess:

- **GAR** reflects the allocation status of resources (i.e., whether GPUs are assigned to jobs);
- **GPU Utilization** reflects the usage intensity (i.e., how actively the GPUs are being used for computation).

For instance, a job may allocate 8 GPUs while experiencing low computational activity (e.g., due to data loading bottlenecks), resulting in high GAR but low GPU utilization. Therefore, both metrics should be analyzed together to achieve a comprehensive understanding of system performance.

## 4.2 Scheduling Occupation Ratio (SOR)

**Definition**

The Scheduling Occupation Ratio (SOR) measures the efficiency of the scheduling system in utilizing GPU resources over time, reflecting the cumulative degree of resource allocation across the cluster. It is defined as:

$$SOR = \frac{Total\ GPU-hours\ allocated\ by\ AI\ workloads}{Total\ available\ GPU-hours\ in\ the\ cluster}$$

**Computation Methodology**

Depending on the job scheduling semantics, GPU-hour accumulation begins upon successful scheduling and resource binding:

- **Gang-scheduled jobs** (e.g., distributed training tasks): GPU-hours are counted at the **job level** once all constituent Pods have been successfully scheduled and GPU resources are reserved;
- **Non-gang-scheduled jobs** (e.g., inference services with traditional non-distributed architectures): GPU-hours are accumulated at the **Pod level**, starting from the moment each Pod is scheduled and GPU resources are allocated.

**Explanation and Analysis**

SOR can be viewed as a time-weighted extension of the GPU Allocation Ratio (GAR), emphasizing the temporal continuity and time-based utilization efficiency of resource usage. Unlike GAR, which captures only the instantaneous allocation state at a given moment, SOR leverages cumulative computation over time to provide a more comprehensive assessment of the scheduling system's long-term resource orchestration capability.

It is important to note that GPU allocation is considered effective from the point of **scheduling completion**—i.e., when the scheduler assigns a Pod to a node and reserves GPU resources—even if the container has not yet entered the Running state. This period (e.g., image pulling, initialization) still consumes allocatable GPU capacity and is therefore included in SOR.

SOR is influenced by the following factors:

- **Scheduling algorithm efficiency**: including job queuing delay, rescheduling latency, and scheduler processing overhead;
- **Platform performance bottlenecks**: such as Pod creation time in Kubernetes, container image pulling duration, and end-to-end latency from scheduling completion to container startup.

Under stable underlying platform performance (e.g., consistent node boot time, network conditions, and image registry latency), optimizing the scheduling algorithm—such as reducing queuing time and improving resource matching efficiency—can effectively improve SOR.

## 4.3 GPU Node Fragmentation Ratio (GFR)

**Definition**

A node is classified as a "non-fragmented node" if it meets either of the following conditions:

- No GPU cards are allocated (completely idle);
- All GPU cards have been allocated (completely occupied).

Otherwise, the node is considered a "fragmented node".

The degree of fragmentation of a fragmented node depends on the usability of its remaining GPUs: the harder the remaining resources are to match mainstream job request patterns, the higher the fragmentation. For simplicity in estimation, the following rule of thumb can be applied: the fewer GPUs allocated and the more remaining, the higher the fragmentation degree.

**Explanation and Analysis**

GPU Node Fragmentation Ratio (GFR) reflects the proportion of partially occupied nodes and is an important indicator of resource allocatability. A higher GFR indicates more nodes are in an intermediate state of "neither idle nor fully loaded", increasing scheduling difficulty.

Note that GFR is a coarse-grained metric that only measures the proportion of "partially occupied" nodes, without distinguishing the actual usability of their remaining resources. For example, in certain LLM training scenarios where jobs typically request 4 or 8 GPUs, nodes with 4 or 6 remaining GPUs may be more usable than those with 1 or 3. Therefore, GFR should be analyzed in conjunction with job size distribution and actual scheduling success rates to more accurately assess a scheduling system's resource consolidation capability.

When GFR remains consistently high, it may indicate resource dispersion due to scheduling strategy limitations. In such cases, introducing a fragmentation consolidation mechanism can improve resource concentration and enhance the acceptance rate and training performance of large jobs.

### 4.4 Job Waiting Time Distribution (JWTD)

**Definition**

Statistics are collected on average waiting time, categorized by job size, including queuing time and scheduling decision-making time.

**Explanation and Analysis**

In AI clusters, job sizes vary significantly. Without categorization, it is difficult to accurately reflect scheduling system performance. For example:

- Small-scale jobs (e.g., requiring fewer than 8 GPUs): face less resource competition and are scheduled quickly;
- Large-scale jobs (e.g., requiring more than 64 GPUs): face intense competition and high scheduling complexity.

Shorter waiting times indicate faster system response and more efficient resource allocation. This metric can be used to evaluate the effectiveness of queue management, preemption mechanisms, and resource admission strategies.

### 4.5 Job Training Time Estimation Distribution (JTTED)

**Definition**

To capture the impact of scheduling strategies on job training performance, this paper introduces "training time estimation" as an evaluation metric. It categorizes statistics by job size for the following two deviation ratios:

- NodeNum Deviation Ratio = Actual allocated node number / Optimal node number
- NodeNetGroupNum Deviation Ratio = Actual cross-group number / Optimal group number

The "optimal node number" refers to the minimum number of nodes that can be allocated while maintaining all-to-all communication within a single LeafGroup, where intra-group bandwidth is maximized and latency is minimized.

**Explanation and Analysis**

Although the impact of scheduling on training performance can be reflected in training duration, the latter is susceptible to various confounding factors such as AI framework-level optimizations. To isolate the effect of scheduling, this paper adopts the "Deviation Ratio" as a proxy metric to more accurately assess how closely a scheduling solution approaches the optimal communication topology. A smaller deviation ratio indicates better alignment with the ideal topology and, consequently, shorter expected training time.

This metric is particularly useful for evaluating scheduling strategies in communication-intensive tasks (e.g., LLM training), offering strong guidance for optimization.

## 4.6 Comprehensive Evaluation Criteria

To holistically evaluate the performance of an AI cluster scheduling system, we propose the following criteria:
Under the same business input conditions, a better scheduling system should exhibit the following characteristics:

1. Higher GAR is preferable, indicating higher GPU utilization efficiency in active resource pools. Under full cluster load, a higher GAR also reflects stronger scheduling capability.
2. Higher SOR is preferable, reflecting sustained high-efficiency resource utilization over time. Under the same load, a higher SOR indicates greater scheduling efficiency per unit time.
3. Lower GPU Node Fragmentation Ratio (GFR) is preferable, indicating stronger resource consolidation and less waste.
4. Shorter Job Waiting Time (JWTD) is preferable, reflecting faster response and higher scheduling efficiency.
5. Lower Job Training Time Estimation (JTTED) is preferable, indicating scheduling closer to optimal communication topology and better training performance.

These five core metrics collectively form a multi-dimensional evaluation framework, effectively supporting comparison, optimization, and evolution of different scheduling strategies.

It is important to note that these metrics may be interrelated. For example:

- High GAR may coexist with long JWTD (high utilization but long queuing);
- Low GFR does not guarantee low JTTED (less fragmentation but more cross-group communication);
- High SOR may result from long job runtime rather than high scheduling efficiency.

Therefore, we recommend a **multi-metric joint analysis approach** in system evaluation to comprehensively assess the strengths and weaknesses of scheduling strategies, avoiding bias from single-metric focus.

## 5 EXPERIMENTAL EVALUATION

This chapter systematically evaluates the scheduling performance of the Kant system in AI clusters of different scales. The experimental environment includes typical large-scale training clusters and small-scale inference clusters, aiming to comprehensively assess its performance in resource utilization, scheduling efficiency, and service quality.

The experiments focus on analyzing the following core metrics: GPU Allocation Rate (GAR), Scheduling Occupancy Rate (SOR), GPU Node Fragmentation Rate (GFR), Job Waiting Time Distribution (JWTD), and Job Training Time Estimation Distribution (JTTED), to support quantitative comparison and optimization of scheduling strategies.

**5.1 Large-Scale Training Cluster**

The experiment was conducted on a homogeneous 8,000-GPU cluster, primarily used for distributed LLM training tasks. Job sizes range from 1 to 2048 GPUs, resulting in intense resource competition and high demands on the scheduling system.

Experimental results show that, compared to the native scheduling system, the Kant system demonstrates superior performance:

- GPU Allocation Rate (GAR) is significantly improved;
- Scheduling Occupancy Rate (SOR) is further enhanced;
- GPU Node Fragmentation Rate (GFR) is noticeably reduced;
- Job Waiting Time (JWTD) is shortened;
- Job Training Time Estimation Deviation (JTTED) is closer to the optimal topology, reducing communication overhead.

The following sections provide an in-depth analysis of these metrics using experimental data.

*5.1.1 Job Characteristics*

Job sizes in the test cluster range from 1 to 2048 GPUs, with most jobs requesting no more than 8 GPUs. However, despite their small number, large jobs consume a disproportionately high share of total GPU resources (Figure 2).

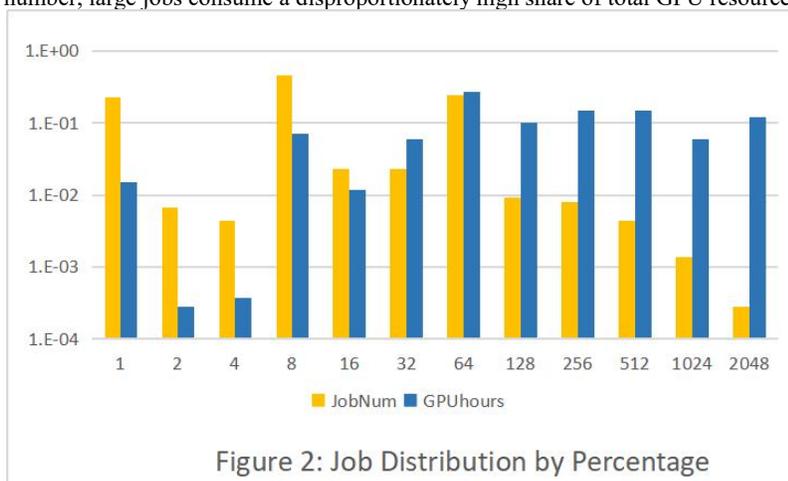

Figure 2: Job distribution by percentage.

*5.1.2 Backfill Experiment*

We use the native scheduling system as the baseline, which typically employs a Strict FIFO mode. In this mode, if the head-of-queue job cannot be scheduled due to insufficient resources, the entire queue becomes blocked.

With Backfill scheduling is enabled in Kant, smaller jobs can bypass the unschedulable head job and utilize idle resources first, improving resource allocation efficiency, as reflected by increased GPU Allocation Ratio (GAR) and Scheduling Occupation Ratio (SOR).

Additionally, when the waiting time of a head-of-queue job reaches a predefined threshold, the system attempts to reallocate resources, and if necessary, preempts them to fulfill the job's requirements. Although preemption incurs additional scheduling and platform overhead—potentially degrading GAR and SOR—the Backfill mode still improves both metrics overall. Experimental data show that, compared to the baseline, the median SOR gain is approximately 3.6%, and GAR remains high with moderate improvement (Figure 3).

Given that the current preemption mechanism still incurs overhead, further optimization of scheduling efficiency and resource release granularity is expected to unlock greater performance potential.

The JWTD indicator remains stable under Backfill (Figure 4), indicating minimal impact on job response latency. Meanwhile, since the initial GFR of the test cluster is already very low (<1%), Backfill has little effect on GFR (Figure 5).

In contrast, under the Best-Effort FIFO mode, although head-of-line blocking is alleviated and relatively high improvements in GAR and SOR are achieved, the lack of preemption causes large jobs to remain resource-starved for extended periods. Results show that waiting times (JWT) for 1024-GPU and 2048-GPU jobs increase significantly (Figure 4). Therefore, Best-Effort is better suited for latency-insensitive workloads.

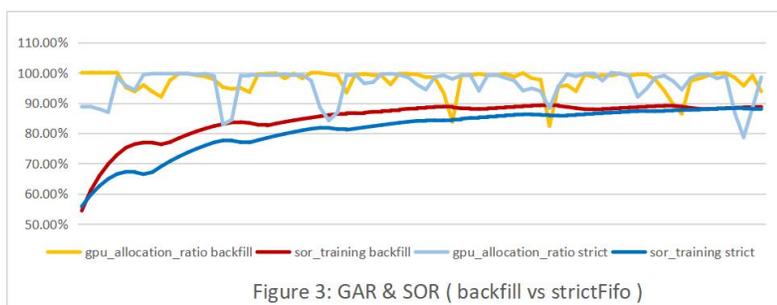

Figure 3: GAR and SOR comparison between Backfill and Strict FIFO.

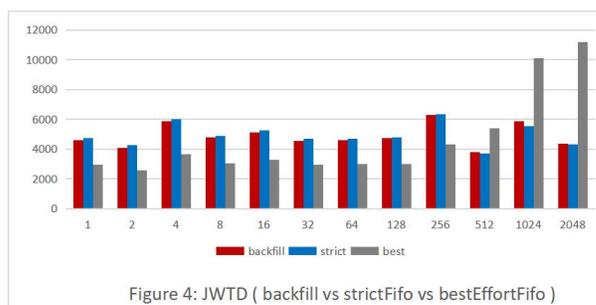

Figure 4: JWTD comparison among Backfill, Strict FIFO and Best-Effort FIFO.

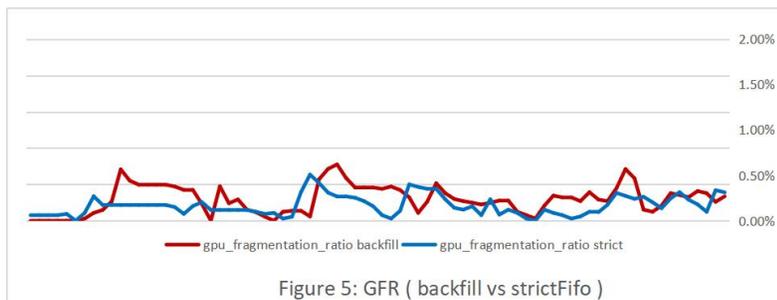

Figure 5: GFR comparison between Backfill and Strict FIFO.

*5.1.3 E-Binpack Experiment*

Using the native scheduling system as baseline, enabling the E-Binpack strategy in Kant significantly reduces GPU node fragmentation rate (GFR), dropping from an average of 8.5% to below 1% (Figure 6).

With reduced fragmentation, both GAR and SOR improve. Experimental data show median gains of approximately 4.1% in SOR and 4.6% in GAR compared to the baseline (Figure 7).

In terms of job response, JWTD shows improvement, with average waiting times decreasing across all job sizes (Figure 8).

Regarding job performance, JTTED also improves: except for 2048-GPU jobs, the estimated average training duration for other job sizes decreases (Figure 9).

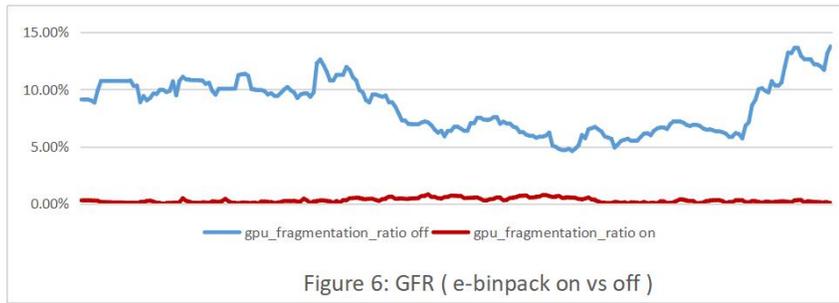

Figure 6: GFR with E-Binpack enabled vs. disabled.

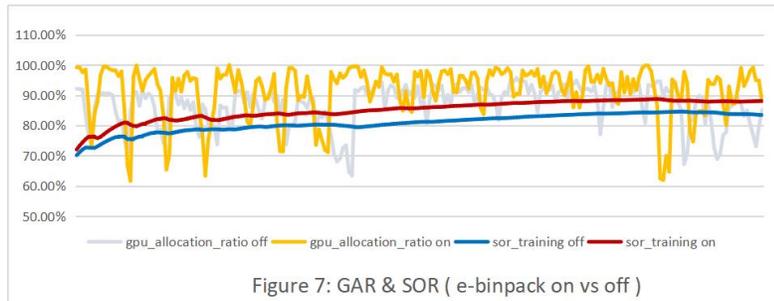

Figure 7: GAR and SOR with E-Binpack enabled vs. disabled.

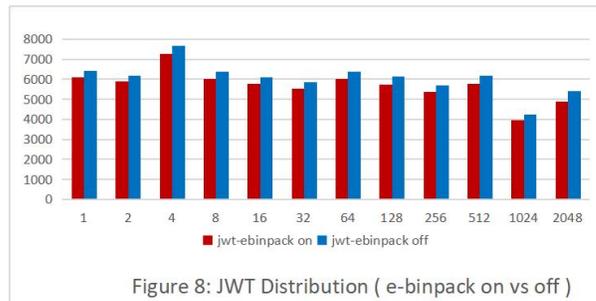

Figure 8: JWTD with E-Binpack enabled vs. disabled.

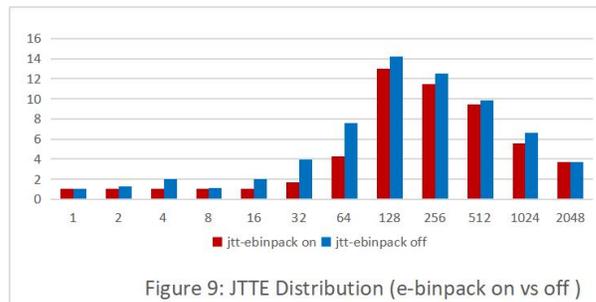

Figure 9: JTTED with E-Binpack enabled vs. disabled.

## 5.2 Small-Scale Inference Cluster

The experiment was conducted on several homogeneous and heterogeneous GPU clusters under the thousand-GPU scale, primarily for multi-tenant inference task scheduling. This scenario has strong real-world representativeness.

Experimental results show that the Kant system maintains strong scheduling performance:

- Supports GPU quota management and fair scheduling in multi-tenant and heterogeneous environments;
- Delivers rapid response and high availability for inference tasks;
- Maintains a high GPU Allocation Rate (GAR);
- Keeps GPU node fragmentation rate (GFR) within a reasonable range.

These results confirm that the Kant system also delivers excellent scheduling performance and QoS in inference scenarios. Specific data are analyzed below.

### 5.2.1 GPU Quota

In a multi-tenant heterogeneous cluster, tenants are allocated varying GPU quotas, and quota utilization varies (Figure 10).

By GPU model, node pool resources are shared among multiple tenants (Figures 11 and 12). Each tenant may also be assigned quotas across multiple GPU models.

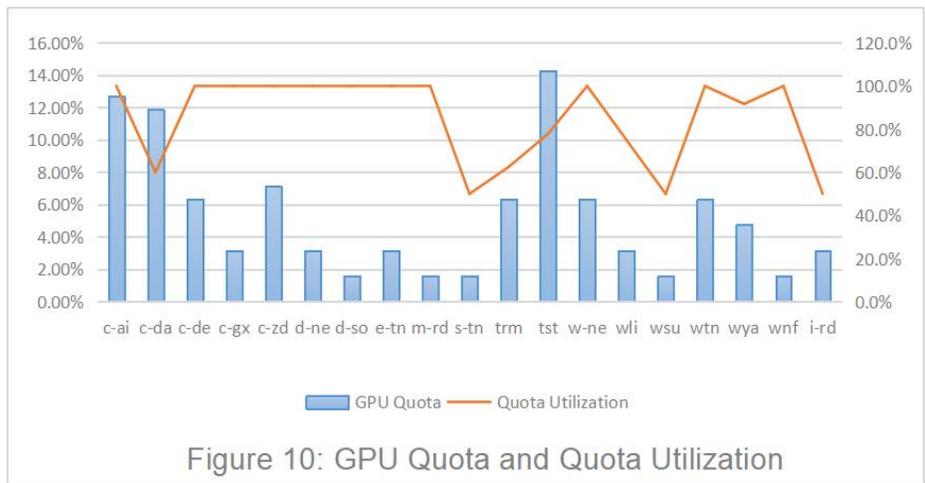

Figure 10: GPU quota and quota utilization.

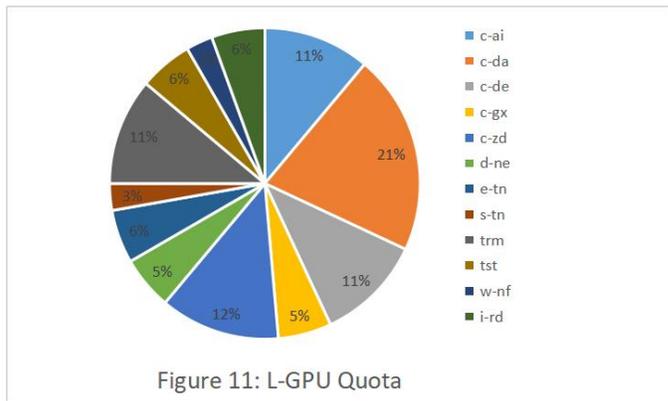

Figure 11: Type-L GPU quota.

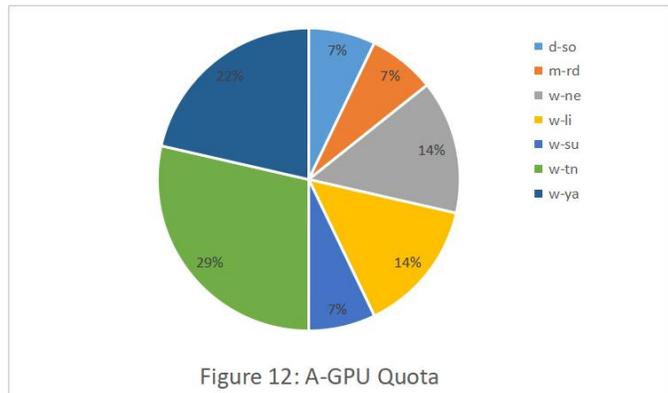

Figure 12: Type-A GPU quota.

*5.2.2 GPU Allocation*

During the observation period in a hundred-GPU heterogeneous inference cluster, job demand approached but did not surpass cluster capacity. No jobs were pending, and GAR remained stable at a high level of approximately 93% (Figure 13). SOR continued to rise and remained high, indicating sustained efficient resource utilization.

GAR fluctuations are primarily driven by job arrivals and completions. Notably, changes in GAR do not necessarily affect GFR—if job changes occur on fully allocated or fully idle nodes, no additional fragmented nodes are created, keeping GFR stable.

The cluster's average GFR was 6.5% (Figure 14). It is important to note that GFR should not be directly compared across clusters of different scales: under the same task change frequency, smaller clusters are more sensitive to individual fragmented nodes, leading to higher GFR (Figure 15, where as cluster size decreases from i7 to a10, GFR increases correspondingly).

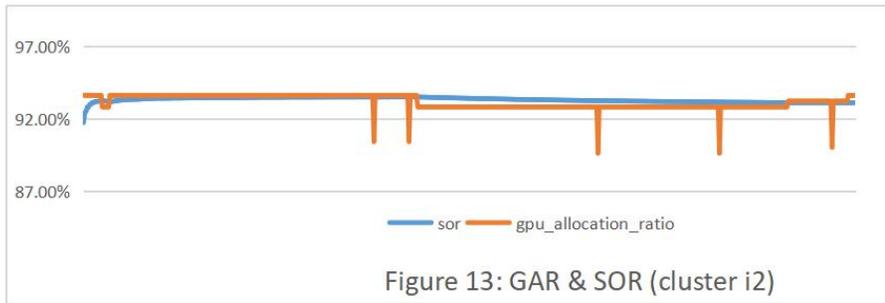

Figure 13: GAR and SOR in cluster i2.

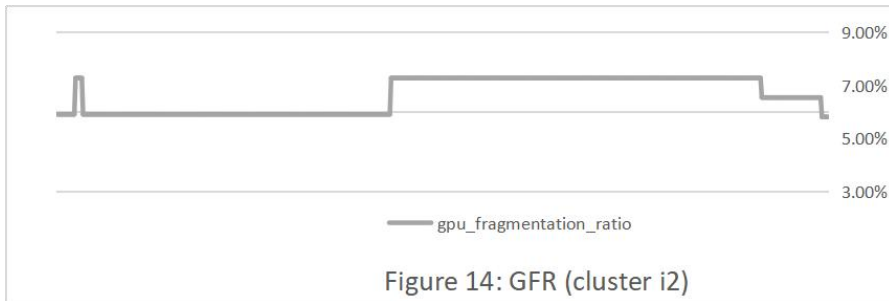

Figure 14: GFR in cluster i2.

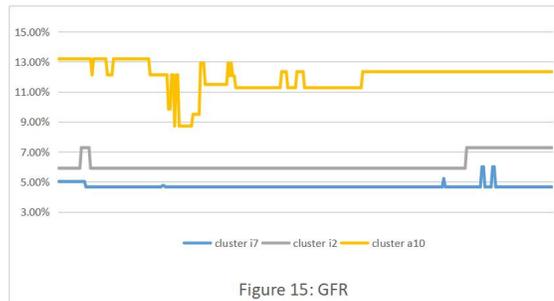

Figure 15: GFR comparison among clusters i7, i2 and a10.

## 6 SUMMARY AND OUTLOOK

This paper presents the Kant system — an efficient scheduling platform for large-scale AI container clusters, supporting unified scheduling for both training and inference jobs.

Based on the practice of the Kant system, this paper systematically organizes and defines a set of key evaluation metrics applicable to AI cluster scheduling, covering multiple dimensions such as resource utilization, scheduling efficiency, and service quality. These metrics not only provide a scientific basis for system performance evaluation but also offer quantitative references for optimizing subsequent scheduling algorithms.

**Experiment Results**

Experimental results demonstrate that the Kant system performs excellently across multiple key scheduling metrics:

1. The GPU Allocation Rate (GAR) and Scheduler Occupancy Rate (SOR) remain at a high level, reflecting efficient resource scheduling.
2. The GPU Node Fragmentation Rate (GFR) is effectively controlled, indicating strong resource integration capabilities.
3. The overall job waiting time (JWTD) remains at a low level, demonstrating the system's strong responsiveness.
4. In training job scheduling, the JTTED is effectively controlled, indicating that communication overhead between nodes and across NodeNetGroups is significantly reduced, thereby demonstrating the system's potential for optimizing training performance.

Currently, the Kant system has been deployed and is operating stably across multiple AI container clusters, reliably supporting large-scale intelligent computing task scheduling requirements.

**Future Work**

In the future, we will continue to advance the functional expansion and performance optimization of the Kant system, with key focus areas including:

1. Continuously improving the system's response speed and scheduling throughput.
2. Enhancing fault-tolerant rescheduling mechanisms for the scheduling system in AI training and inference scenarios to ensure high availability and operational stability.
3. Exploring cross-cluster and cross-regional joint scheduling capabilities to build a unified global resource view and coordinated scheduling framework.
4. Deepening collaborative optimization with AI training frameworks and inference engines to comprehensively enhance the computing efficiency of AI clusters.

Our long-term goal is to closely follow the development trends of AI business and build an AI cluster scheduling platform that covers diverse AI application scenarios and combines high performance with intelligent capabilities. Through continuous technological innovation and engineering practice, we are committed to establishing an efficient, flexible, and scalable underlying scheduling infrastructure for next-generation AI systems, continuously advancing the development of AI cluster scheduling technologies.